\documentclass[12pt,preprint]{aastex}

\shorttitle{ Mid-IR Interferometry of RS~CrB at Keck }
\shortauthors{B. Mennesson et al.}

\begin{document}


\title{ The dusty AGB star RS~CrB: first mid-infrared interferometric 
observations with the Keck Telescopes }


\author{B. Mennesson\altaffilmark{1}, C. Koresko\altaffilmark{2}, M.J. 
Creech-Eakman\altaffilmark{3}, E. Serabyn\altaffilmark{1}, M.M. Colavita
\altaffilmark{1}, R. Akeson\altaffilmark{2}, E. Appleby\altaffilmark{4}, J. 
Bell\altaffilmark{4}, A. Booth\altaffilmark{1}, S. Crawford\altaffilmark{1},
W. Dahl\altaffilmark{4}, J. Fanson\altaffilmark{1}, C. Felizardo
\altaffilmark{2}, J. Garcia\altaffilmark{1}, J. Gathright\altaffilmark{4}, J. 
Herstein\altaffilmark{2}, E. Hovland\altaffilmark{1}, M. Hrynevych\altaffilmark{4},
E. Johansson\altaffilmark{4}, D. Le Mignant\altaffilmark{4}, 
R. Ligon\altaffilmark{1}, R. Millan-Gabet\altaffilmark{2}, J. Moore
\altaffilmark{1}, C. Neyman\altaffilmark{4}, D. Palmer\altaffilmark{1}, T. 
Panteleeva\altaffilmark{4}, C. Paine\altaffilmark{1}, S. Ragland\altaffilmark{4},
L. Reder\altaffilmark{1}, A. Rudeen\altaffilmark{4}, T. Saloga
\altaffilmark{4}, M. Shao\altaffilmark{1},  R. Smythe\altaffilmark{1}, K. Summers
\altaffilmark{4}, M. Swain\altaffilmark{1},  K. Tsubota\altaffilmark{4},
C. Tyau\altaffilmark{4}, G. Vasisht\altaffilmark{1}, P. Wizinowich
\altaffilmark{4}, J. Woillez\altaffilmark{4}}


\altaffiltext{1}{Jet Propulsion Laboratory, California Institute of 
Technology, 4800 Oak Grove Drive, Pasadena CA 91109-8099}
\altaffiltext{2}{Michelson Science Center, California Institute of 
Technology, 770 South Wilson Avenue, Pasadena CA 91125}
\altaffiltext{3}{New Mexico Tech, Dept of Physics, 801 Leroy Place,
Socorro, NM 87801}
\altaffiltext{4}{W.M. Keck Observatory, California Association for Research
in Astronomy, 65-1120 Mamalahoa Highway, Kamuela, HI 96743 }

\begin{abstract}

We report interferometric observations of the semi-regular variable star 
RS~CrB, a red giant with strong silicate emission features.  The data were
among the first long baseline mid-infrared stellar fringes obtained between
the Keck telescopes, using parts of the new nulling beam combiner.  
The light was dispersed by a low-resolution spectrometer, allowing simultaneous
measurement of the source visibility and intensity spectra from 8--12 $\mu$m.
The interferometric observations allow a non-ambiguous determination
of the dust shell spatial scale and relative flux contribution.
Using a simple spherically-symmetric model, in which a geometrically 
thin shell surrounds the stellar photosphere, we find that $\sim$ 30\% to
 $\sim$ 70\% of the overall mid-infrared flux - depending on the wavelength -
originates from 7-8 stellar radii. The derived shell opacity
profile shows a broad peak around 11 microns ($\tau \simeq
$ 0.06), characteristic of Mg-rich silicate dust particles.

\end{abstract}


\section{Introduction}

The apparent interferometric sizes of variable red giants vary dramatically - 
up to a factor of 3 - with wavelength and pulsation phase \cite{tuthill95,
tuthillspie,weiner,mennesson2002} \\  \cite{thompson,perrin2004,weiner04}.
These large variations challenge current hydrodynamical and line 
opacity models of these stars, indicating that their extended atmospheres
 are extremely complex. Simultaneous spectroscopic and high angular resolution
observations over a wide wavelength range are key to understanding
the various phenomena these stars exhibit, including photometric
 pulsation, high mass-loss rates, dust nucleation, and formation of molecular
layers and masers.  High angular resolution thermal infrared observations
of bright red giants in {\it narrow bandpasses} ($\Delta\lambda/\lambda
\simeq 10^{-4}- 10^{-3}$) have been available for some time 
\cite{danchi}. The Keck Interferometer Nuller (KIN) \cite{serabyn2004} and 
the MIDI instrument on the VLTI \cite{leinert} offer much greater spectral
 coverage, typically over the whole 8--13 $\mu$m region, with a sensitivity
goal of about 1 Jy. 

The Keck Interferometer is nearly ideal for the study of such stars.
Its 85 m baseline produces a fringe spacing of 27 mas at 11~$\mu$m, 
commensurate with the expected angular scales of most variable red
 giants located within 1\,kpc (photosphere diameter of a few mas, with 
thermal emission from dust peaking within several stellar radii).  The 
detection is carried out by means of a low-resolution spectrometer ($\lambda/\Delta\lambda
\simeq$ 35), sensitive across a bandpass which spans the
 broad opacity peaks characteristic of various types of O-rich and C-rich
dusts.

RS~CrB was the first star successfully observed with the Keck Interferometer's
mid-infrared (MIR) beam combiner.  
It is a bright semi-regular variable of type {\it a}, quite similar to a 
Mira but with smaller photometric variations ($\Delta m_{V}$ $\simeq$ 1-2).
It has a mean pulsation period of 332 days \cite{kholopov} and is 
classified as a red giant of spectral type M7, with a 12 $\mu$m flux of 53\,Jy
\citep{iras}. It has an oxygen-rich dust shell with strong silicate emission
features detected by the IRAS Low Resolution Spectrometer (LRS) (class
SE6t, \citep{sloan98}).  
Ita et al. \cite{ita} detected SiO maser emission around 
RS~CrB,  which can be regarded as a tracer of dense molecular gas in the 
inner region ($\simeq$ 2 stellar radii) of the circumstellar envelope (e.g.
VLBA images of TX Cam \cite{diamond} and other Miras \cite{cotton}). 

We present a measurement of the wavelength-dependent fringe visibility
of RS~CrB from 8--12~$\mu$m, a spectral region where both the central
star and its surrounding dust shell are visible. 
In combination with the intensity spectrum measured across the same band 
with the individual Keck telescopes, it is possible to determine the relative
brightness of the two components, the spatial extent of the shell, and
its opacity spectrum.



\section{ Observations}


The observations were made on August 8 2004, using a subset of the KIN optics
\citep{mennesson2003,serabyn2004} as a two-input  beam combiner, without
most of the special features needed to produce the high fringe contrast
($>$ 0.99) required for nulling.
The telescopes each fed one of the two inputs of the primary modified Mach-
Zehnder (MMZ) symmetric beam combiner \citep{serabyn2002}.  One of the 
two resulting outputs was sent to 
KALI, a low-resolution MIR spectrometer \citep{mce2003}. KALI uses
a Boeing Si:As BIB detector, cooled to 4\,K, with the optics and radiation
shield cooled to $\sim 90$~K using LN$_2$.   A cold internal pinhole
(equivalent diameter 2$\lambda$/D) at an intermediate focus provides spatial
filtering. Direct-view prisms disperse the N-band light into 14 spectral
channels.
The central wavelengths for the channels were measured using Fourier Transform
Spectroscopy (FTS) scans of a hot filament.  They ranged
from 8.16--12.01 $\mu$m, with the channel width varying from $\simeq$
0.2 $\mu$m on the red end to $\simeq$ 0.4 $\mu$m on the blue end.

\begin{figure}
\epsscale{.50}
\plotone{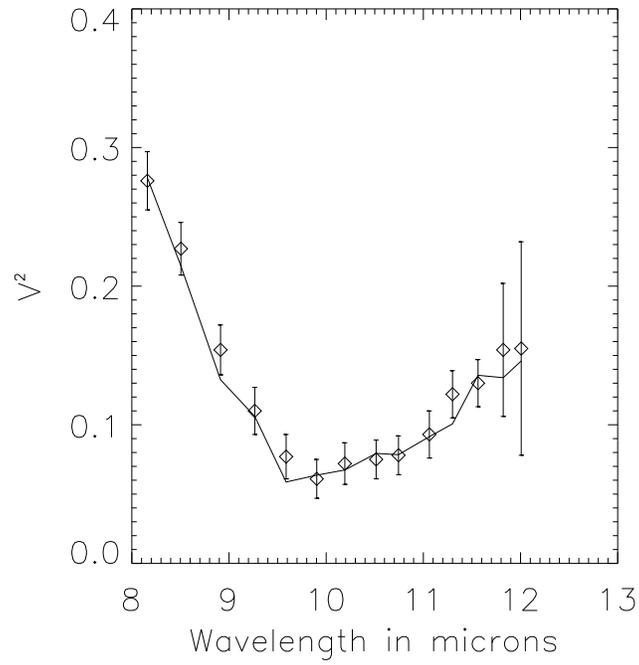}
\caption{Diamonds and error bars (1-$\sigma$): observed visibility spectrum
for RS~CrB. Plain curve: best model fit.\label{v2}}
\end{figure}

RS~CrB was observed at a single, quasi constant projected baseline (80.6--
80.9\,m), along with a set of calibrator stars (early K giants) with
known near-infrared (NIR) diameters. To allow for the differential limb darkening
between the NIR and MIR wavelengths,
we adopted conservative error bars on the estimated calibrators diameters
in the MIR. A list of the stars and their properties is given 
in Table \ref{starlist}.  The observation sequences were 
similar to those used at NIR wavelengths for the Keck interferometer
\citep{colavita2003}.  The fringe amplitude was measured accurately
 through a fast (40\,Hz) optical path difference modulation.
Each star was visited between 3 and 4 times, with typically one or two 
minutes spent tracking the fringe during a visit.
The most important source of uncertainty in these visibility measurements
was  the 
removal of the  strong and variable thermal background. For each of the 
two beams, the absolute stellar signal  was determined through sky chopping.
A spatial chop was implemented using the fast tip-tilt mirrors of the 
Keck adaptive optics systems \cite{wizi2003} at a frequency of  5\,Hz. 
Because of beamwalk on the optics upstream of the tip-tilt mirror, a 
fraction of the thermal background is also modulated. To correct for this 
bias, the demodulated background signal was determined through a similar 
chopping measurement on blank sky. After this calibration, stellar
signals derived from the various single aperture observations still exhibited
fluctuations at the $\simeq$ 5-10\% level, which we hope to overcome 
in the future with a symmetric three position chop. This is currently the 
main contributor to the visibility error bars, both on RS CrB and the 
calibrators. The data reduction was otherwise similar to that used for NIR 
visibility measurements at PTI and KI \cite{colavita1999}.  The calibrator
observations were used to measure the ``system visibility'' (the response
of the interferometer to a point-source), which did not show any 
significant time dependence.
Its value during the RS~CrB observations was estimated by combining 
the system visibilities measured on all three calibrators (Table 1).  The
 resulting calibrated visibility curve for RS~CrB is plotted in Fig \ref{v2},
showing that the source is clearly resolved.

\section{Modeling and interpretation}

Since all measurements presented here were obtained with a single baseline
orientation, only models with spherical symmetry are considered.  There
is also astrophysical justification for this choice, both on a 
theoretical and observational basis. 
The envelope surrounding an AGB star can be described to first order by a
spherically symmetric outflow at a constant velocity of $\sim 10$ km/sec,
in which the material is heated by friction between gas and dust particles
and cooled primarily by the line radiation of molecules such as H$_2$O
\cite{habing}.
Also, maser observations of {\it single} evolved stars generally show 
ring-like structures which are much more compatible with spherical geometries
than disk-like ones \cite{diamond,cotton}. 
Finally, recent NIR interferometric phase closure measurements
 failed to detect asymmetric structures in a sample of 10 M-class 
semi-regular variables (S. Ragland, priv. comm.).  

\begin{figure}
\epsscale{.50}
\plotone{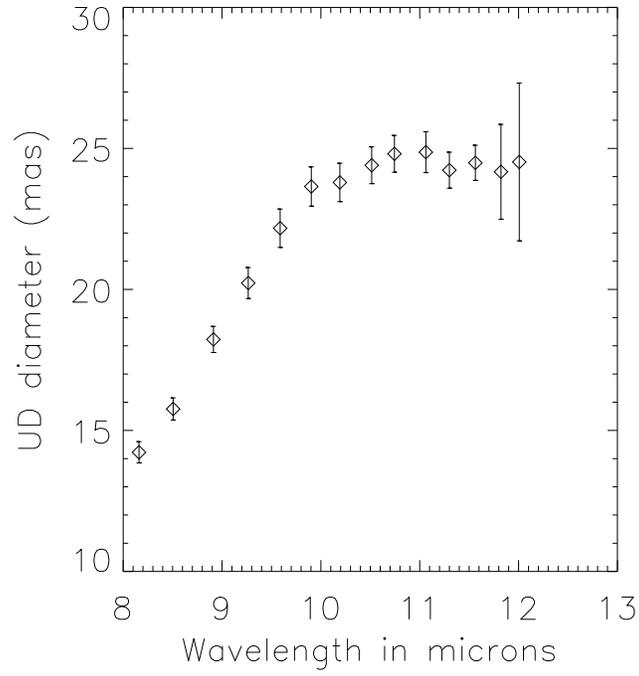}
\caption{Uniform disk diameters obtained when fitting the observed $V^{2}$
at each wavelength.  \label{UD_vs_wavelength}}
\end{figure}

We do not expect single component brightness distributions to accurately 
describe dusty red giants in the MIR. Yet, a uniform disk fit
allows a simple representation of the visibility data at each wavelength,
and we use it here as a convenient starting point.   
The resulting apparent diameter, plotted in Fig.~\ref{UD_vs_wavelength},
increases rapidly between 8.2 and 10 $\mu$m and reaches a plateau at longer
wavelengths. The apparent size and its chromatic variations are 
large, pointing to a variable opacity effect in the upper atmosphere, rather
than to true changes in the physical dimension of a single photosphere.
Adopting now a two component representation, consisting of a central 
photosphere surrounded by a dust shell with a variable opacity, the visibility
is sensitive to the size of the shell and its brightness relative to 
the central star, and depends only weakly on the size of the stellar 
photosphere.
 A straightforward interpretation to the apparent size variations in 
Fig.~\ref{UD_vs_wavelength} is that the relative brightness of the shell
increases with wavelength across the measured band, becoming dominant at
the long wavelengths.  This behavior correlates nicely with the SE6 
classification of RS~CrB in the silicate dust sequence \citep{sloan98}, which
 shows a similar variation of the infrared excess versus wavelength.

\begin{figure}
\epsscale{.50}
\plotone{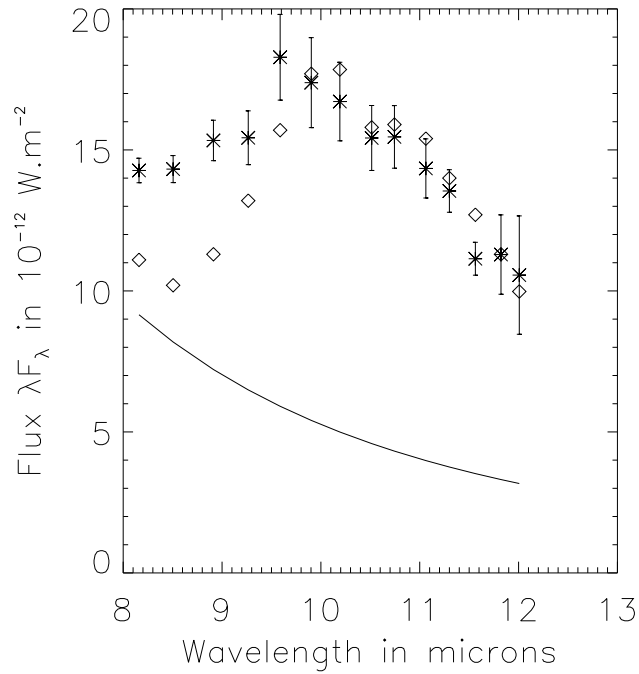}
\caption{RS CrB spectra. Crosses and error bars: observed spectrum using 
the individual Keck telescopes. The raw spectrum was normalized by the KALI
responsivity, 
which is 
determined by comparing raw KALI spectra of the calibrator stars with their IRAS LRS spectra.  
 Diamonds: IRAS LRS spectrum converted to KIN's spectral resolution. The 
Keck measured spectrum has an overall flux close to that measured two decades
earlier by IRAS, 
but shows significantly more flux at the blue end.
Plain curve: naked star spectrum, for the best $V^{2}$ fit parameters. 
\label{photometry_plot}}
\end{figure}

We now model the shell as infinitely thin, with angular diameter $\Phi_{shell}$,
optical depth $\tau(\lambda)$  and temperature $T_{shell}$, surrounding
a blackbody photosphere of angular diameter $\Phi_{*}$ and temperature
$T_{*}$.  This model is similar to the one previously applied to molecular
layers around red giants \citep{perrin2004,scholz2001}.  It predicts
both the visibility spectrum and the intensity spectrum across the 
observed bandpass. We explored a grid of model parameters \{$T_{*}$,  $\Phi_{*}$, $T_{shell}
$, $\Phi_{shell}$ \}, computing $\chi2$ per spectral channel for the 
visibility at each point in the grid.    
The spectrum of  RS~CrB derived from our single Keck telescope observations 
(Fig.~\ref{photometry_plot})  is used to determine $\tau(\lambda)$
at each point.  
We constrain the stellar temperature $T_{*}$ to be 3100 $\pm$ 150\,K, 
corresponding to the range of effective temperatures proposed for M7 giants 
\cite{dyck, perrin1998}.   The value of $\Phi_{*}$ can be estimated
from a blackbody fit to the NIR flux \citep{kersch}.
The results range from 3.7--4.2 mas between the J and M bands.  This 
estimate is likely crude, however, as it neglects the effect of the 
circumstellar envelope. 
In light of this uncertainty, we allowed $\Phi_{*}$ to range from 2.5--5 mas.
The shell diameter was allowed to vary between 1 and 50 times $\Phi_{*}$,
while the maximum shell temperature was set to 1500\,K, a reasonable
upper limit for the grain sublimation temperature. 
The error bar on each parameter is computed as the deviation that increases
the $\chi^{2}$  by one after reoptimizing all the other parameters. The
resulting values are:
$\Phi_{*}$ = $3.78 \pm 0.20$\,mas,
$\Phi_{shell}$ = $27.6 \pm 1.2$\,mas,
$T_{shell}$ = $1160 \pm 300$\,K, and
$\tau_{max}=\tau_{11.1\mu m}$ = 0.04-0.3.
The agreement between the model and the observed $V^{2}$ is excellent 
($\chi^{2}$ per point = 0.38, Fig.~\ref{v2}, plain curve) showing that the
 visibility data at all wavelengths can be reproduced by a 2-component 
model with a fixed, {\it wavelength independent} geometry, 
and an opacity profile fixed by the observed intensity spectrum. 
The shell diameter $\Phi_{shell}$ is the most accurately constrained 
parameter.  
It indicates that most of the flux at long wavelengths - probing the dust
emission zone- originates from a region located around 7-8 stellar radii.
Interestingly, this is comparable to the inner dust shell size derived
by the MIDI instrument for the Mira star RR Sco \citep{rrsco}.
For wavelengths shorter than 10 $\mu$m, the contribution from the central
star is visible and well constrained by the observed $V^{2}$. For the 
stellar temperature range adopted, the fit yields a central object diameter
consistent with the NIR photometry.
We expect the photospheric diameter as measured at shorter NIR wavelengths to be smaller, due to the effect
of molecular layers around this type of star \cite{hinkle,tsuji,yamamura}
which is more prominent at long infrared wavelengths \cite{mennesson2002,
weiner04}. This has now been confirmed by IOTA interferometric observations 
at 1.65$\mu$m - a region free of significant molecular absorption -,
which yield an uniform disk diameter of 3.15 $\pm$ 0.12 mas for RSCrB
(John Monnier, priv. comm., June 2005).

\begin{figure}
\epsscale{.50}
\plotone{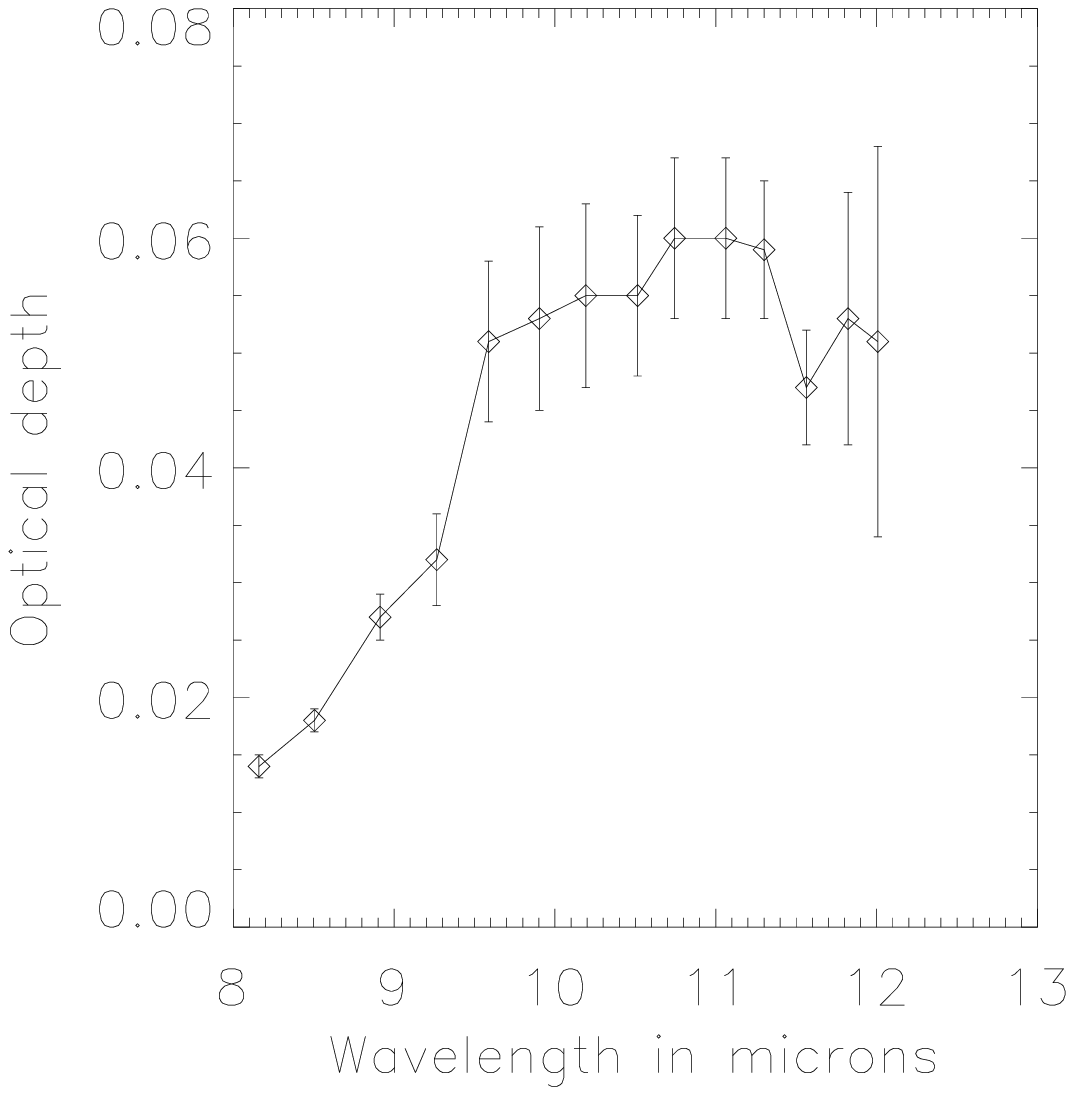}
\caption {Optical depth profile derived from the single telescope spectra,
using the best model parameters: $\Phi_{*}$ = 3.78\,mas, $T_{*}$ =
3100\,K, $\Phi_{shell}$ = 27.6\,mas and $T_{shell}$ = 1160\,K. 
\label{tau}}
\end{figure}

The model fit depends on the overall flux from the shell rather than on 
$T_{shell}$ and $\tau(\lambda)$ individually, so those parameters are not 
separately well constrained by the fit.  $T_{shell}$ must be at least
$\sim 850$\,K to account for the observed shell flux at the shortest 
wavelengths, and is presumably below $\sim 1450$\,K, as that is the maximum
radiative equilibrium temperature allowing silicate grains to survive 
in a stationary flow around a typical mass-losing AGB star 
\citep{willson}.  Within this temperature range, the corresponding 
$\tau(\lambda)$ varies over almost an order of magnitude. 
Although the overall dust opacity is not well constrained, its {\it shape}
is (Fig.~\ref{tau}). The sharp opacity rise observed between 8 and 9.7 
$\mu$m is characteristic of amorphous silicates. The shoulder seen near 
11 $\mu$m is often seen in low mass-loss rate AGB stars \cite{waters} and may
indicate the presence of another solid, such as amorphous alumina. It 
could for instance be in the form of Mg-Fe aluminosilicate compounds 
\cite{mutschke}.  Alternatively, this 11 $\mu$m feature could reflect the 
presence of Mg-rich crystalline forms of olivine such as 
forsterite ($Mg_{2}SiO_{4}$), which exhibits an absorption peak around 
11.2$\mu$m \cite{koike}. Our spectral resolution does not allow to choose 
between these two different scenarios at present.

\section{Conclusions}

Using broad-band thermal interferometry with the twin Keck telescopes, we
 have measured the size and opacity profile of the dust shell around RS~CrB,
an AGB star with strong silicate emission features. The dust 
shell ``radius'' derived in our model can  be understood  as a  characteristic size of the
 weighted brightness distribution of the shell. Strikingly, a single shell radius value of 7-8 $R_{*}$ allows to fit the data over the whole 8-12 microns range, suggesting that most of the detected emission arises from dust in this region. The opacity profile which fits both 
the intensity and  visibility spectra indicates the presence of amorphous silicates, and possibly other  dust species like amorphous alumina or Mg-rich olivine in this region. The $V^{2}$ measurements obtained at the blue end of our bandpass, where 
the dust opacity is low, constrain the size and fractional amount  of flux
coming from the ``central'' star. 
These first results obtained with the Keck Interferometer already demonstrate the value 
of wide band long baseline thermal infrared interferometry. Interferometric
observations allow non ambiguous determination of the spatial brightness
distribution very close to the central source, information that cannot
 be derived from spectroscopy alone. Finally, the wide spectral coverage 
provides access to different physical regimes and can be used to 
simultaneously constrain multiple components of an astronomical source.








The Keck Interferometer is funded by the National Aeronautics and Space 
Administration (NASA). Part of this work was performed at the Jet Propulsion
Laboratory, California Institute of Technology, and at the Michelson 
Science Center (MSC), under contract with NASA. The Keck Observatory was 
made possible through the generous financial support of the W.M. Keck 
Foundation. Finally, the authors wish to recognize and acknowledge the very 
significant cultural role and reverence that the summit of Mauna Kea has 
always had within the indigenous Hawaiian community.
We are most fortunate to have the ability to conduct observations from 
this mountain.

\clearpage

\begin{deluxetable}{ccccc}
\tablecolumns{8}
\tablewidth{0pc}
\tablecaption{Target and Calibrator Stars \label{starlist}}
\tablehead{
\colhead{Star} & \colhead{Type} & \colhead{$\theta_*$ (mas)} & \colhead
{Fringe Time (sec)} & \colhead{$V2_{sys}$}
}
\startdata
RS~CrB & M7 SRa & $\sim$4 & 411 &  \\
$\epsilon$ CrB & K2III & 2.81 $\pm$ 0.1 $^{a}$ & 676 & $0.57 \pm 0.15$ \\

39 Cyg & K3III & 2.82 $\pm$ 0.1 $^{a}$ & 604 & $0.67 \pm 0.11$ \\
$\epsilon$ Cyg & K0III & 4.47 $\pm$ 0.2 $^{a,b}$ & 598 & $0.71 \pm 0.08$ \\
\enddata \\
\tablecomments{For RS CrB, the quoted diameter is the one expected from near 
infrared photometry, assuming a spherical blackbody at 3100\,K. The calibrator diameter values are
taken from (a): the CHARM Catalog 
(Richichi et al. 2005), and (b): the Catalog of LBSI Calibrator Stars (Borde et al.
2002).  The $V2_{sys}$ for each calibrator is an average of the values
over the bandpass and over visits to the star.}
\end{deluxetable}




\clearpage

\clearpage

\clearpage

\clearpage

\clearpage




\clearpage

\begin{thebibliography}{}


\bibitem[Beichman et al. 1988]{iras} Beichman C.A. et al., IRAS catalogs 
and Atlases, Version 2, Explanatory Supplement, NASA Publication 1190, I.

\bibitem[Beichman et al. 1999]{beichman1999} Beichman C.A., Woolf, N.J.  
and Lindensmith C.A., 1999, The Terrestrial Planet Finder: a NASA Origins
 Program to Search for Habitable Planets, JPL Publication 99-3.
\bibitem[Colavita et al. 1999]{colavita1999} Colavita, M.M. et al. 1999, 
ApJ 510, 505
\bibitem[Colavita et al. 2003]{colavita2003} Colavita M.M. et al. 2003, 
ApJ, 592, 83
\bibitem[Cotton et al. 2004]{cotton} Cotton W. et al. 2004, A\&A, 414, 275
\bibitem[Creech-Eakman et al. 2003]{mce2003} Creech-Eakman M.J. et al. 2003,
in SPIE Conf. Proceedings 4841, 330
\bibitem[Danchi et al. 1994]{danchi} Danchi W. et al. 1994, AJ, 107, 1469
\bibitem[Diamond and Kemball 1999]{diamond} Diamond P.J. \& Kemball A.J. 
1999, IAU Symposium 191, edited by T. Le Bertre, A. Lebre and C. Waelkens,
195
\bibitem[Dyck et al. 1996]{dyck} Dyck H.M. et al. 1996, 111, 1705
\bibitem[Fogel and Leung 1998]{fogel} Fogel M.E. \& Leung C.M. 1998, ApJ,
 501, 175.
\bibitem[Habing 1996]{habing} Habing H.J. 1996, A\&A Rev. 7, 2, 97
\bibitem[Hinkle \& Barnes 1979]{hinkle} Hinkle K.H. \& Barnes T.G. 1979, 
ApJ, 227, 923
\bibitem[2001]{ita} Ita Y. et al. 2001, A\&A, 376, 112
\bibitem[Kerschbaum and Hron 1994]{kersch} Kerschbaum F. \& Hron J. 1994,
 A\&AS, 106, 397
\bibitem[Kholopov et al. 1992]{kholopov} Kholopov et al. 1992, Bulletin 
d'Information Centre Donnees Stellaires, Vol 40, 15
\bibitem[Koike et al. 2003]{koike} Koike C. et al. 2003, A\&A 399, 1101
\bibitem[Koresko et al. 2003] {koresko2003}  Koresko C.D., Mennesson B.P.
, Serabyn E., Colavita M., Akeson R., Swain M. 2003, SPIE 4838, 625
\bibitem[Leinert et al. 2003]{leinert} Leinert C. et al. 2003, Astrophysics
\& Space Science, 286, 1, 73
\bibitem[Mennesson et al. 2002]{mennesson2002} Mennesson B. et al. 2002, 
\apj, 579, 446
\bibitem[Mennesson et al. 2003]{mennesson2003} Mennesson B. et al. 2003, 
in Conf. Towards Other Earths: DARWIN/TPF and the Search for Extrasolar 
Terrestrial Planets, ed. M. Fridlund \& T. Henning, (Noordwijk: ESA-SP-539),
525.
\bibitem[Monnier et al. 2004]{monnier} Monnier J. et al. 2004, \apj, 605, 436
\bibitem[Mutschke et al. 1998]{mutschke} Mutschke H. et al. 1998, A\&A 333,
188
\bibitem[Ohnaka et al. 2005]{rrsco} Ohnaka K. et al 2005, A\&A, 429, 1057
\bibitem[Perrin et al. 1998]{perrin1998} Perrin G. et al. 1998, A\&A, 331,
619
\bibitem[Perrin et al. 2004]{perrin2004} Perrin G. et al. 2004, A\&A, 426,
279
\bibitem[Scholz  2001]{scholz2001} Scholz M. 2001, \mnras, 321, 347
\bibitem[Serabyn \& Colavita 2001]{serabyn2002} Serabyn E. and Colavita M.
M. 2001, Applied Optics, 40, 1668
\bibitem[Serabyn et al. 2004]{serabyn2004} Serabyn E. et al. 2004, SPIE 
Conf. Ser. 5491, 806, ed. W.A. Traub
\bibitem[Sloan \& Price 1998]{sloan98} Sloan G.C. \& Price S.D. 1998, 
\apjs, 119, 141
\bibitem[Thompson,~ Creech-Eakman,~ \& van Belle 2002]{thompson} Thompson
 R.R., Creech-Eakman M.J. \& van Belle G.T. 2002, ApJ, 124, 1706
\bibitem[Tsuji 1988]{tsuji} Tsuji T. 1988, A\&A, 197, 185
\bibitem[Tuthill,~ Haniff,~ \& Baldwin 1995]{tuthill95} Tuthill, P.G., Haniff,
C.A. \& Baldwin, J.E., 1995, \mnras, 277, 1541
\bibitem[Tuthill,~ Monnier,~ \&  Danchi  2000]{tuthillspie} Tuthill, P.G.,
Monnier, J.D. \&  Danchi, W.C., 2000, SPIE Conf. Ser. 4006, 491,
, ed. P.J. L\'ena \& A. Quirrenbach 
\bibitem[Waters et al. 1998]{waters} Waters R. et al. 1998, A\&A, 331, L61
\bibitem[Weiner et al. 2000]{weiner} Weiner J. et al.  2000, \apj, 544, 1097
\bibitem[Weiner et al. 2004]{weiner04} Weiner J. 2004, \apj, 611, 37
\bibitem[Willson 2000]{willson} Willson L.A. 2000, ARA\&A, 38, 573
\bibitem[Wizinowich et al. 2003]{wizi2003} Wizinowich P. et al. 2003, SPIE
Proc. 4839, 9
\bibitem[Yamamura et al. 1999]{yamamura} Yamamura, I., de Jong T.,\& Cami
 J., 1999, A\&A, 348, L55


\end{thebibliography}
\end{document}